\documentclass[twocolumn,showpacs,preprintnumbers,amsmath,amssymb,aps,pre]{revtex4}

\usepackage{mathtools,amssymb,lipsum}

\usepackage{epsfig}

\usepackage{graphicx,latexsym,xcolor}

\usepackage{tikz}

\usepackage[english]{babel}
\usepackage{amsmath}
\usepackage{color}
\usepackage{epsfig}
\usepackage{graphicx}
\usepackage{bm}
\usepackage{times,amsmath,amssymb}
\usepackage{subfigure}
\usepackage{amsfonts}
\usepackage{amssymb}
\usepackage{color}
\global\long\def\bege{\begin{equation}}
\global\long\def\ende{\end{equation}}

\global\long\def\begal{\begin{align}}
\global\long\def\endal{\end{align}}

\begin{document}

\title{Violation of Bell's inequalities in uniform random graphs}
\author{Ioannis Kleftogiannis$^1$, Ilias Amanatidis$^{2}$}
\affiliation{$^1$ Physics Division, National Center for Theoretical Sciences, Hsinchu 30013, Taiwan }
\affiliation{$^2$Department of Physics, Ben-Gurion University of the Negev, Beer-Sheva 84105, Israel}

\date{\today}
\begin{abstract}
We demonstrate that quantum correlations can emerge from the statistical correlations of random discrete models, without an a priori assumption that the random models are quantum mechanical in nature, that is without considering superpositions of the random structures. We investigate the correlations between the number of neighbors(degree) for pairs of vertices in Erdos-Renyi uniform random graphs. We use the joint probabilities for the appearance of degree numbers between the vertices in the pairs, in order to calculate the respective Bell's inequalities. We find that the inequalities are violated for sparse random graphs with ratio of edges over vertices $R<2$, signifying the emergence of quantum correlations for these random structures. The quantum correlations persist independently of the graph size or the geodesic distance between the correlated vertices. For $R>2$, as the graph becomes denser by adding more edges between its vertices, the  Bell's inequalities are satisfied and the quantum correlations disappear. Relations to our previous works concerning the emergence of spacetime and its geometrical properties from uniform random graphs, are also briefly discussed.
\end{abstract}

\maketitle
The origin of correlations in quantum mechanical systems
has been a fundamental topic of research since the founding
of quantum mechanics. Issues like non-locality of the correlations and conflicts with classical theories has been pointed out in various works, such as the EPR paradigm\cite{epr1}. The ideas were later put formally in rigid mathematical form by John Bell, demonstrating that certain statistically correlated models involving hidden
variables used to interpret the quantum mechanical results, have to satisfy certain inequalities\cite{bell1,bell2,espagnat}. Since then it has been shown several times experimentally\cite{aspect1,aspect2,fine,bellexp1,bellexp2,bellderivWignerform} that quantum mechanical systems violate the Bell's inequalities.

Typically the Bell's inequalities are defined between two spatially separated particles, by measuring the state of each particle and then forming an inequality containing linear combinations of the joint probabilities that the two particles lie at specific combination of states.
The inequality is satisfied by classical statistical models containing hidden variables that correlate the properties of the two particles.
On the other hand, when the inequality is violated quantum mechanical correlations between the two spatially separated particles are implied. 
For our study we consider Erdos-Renyi uniform random graphs $G(n,m)$ with a fixed number of vertices n and edges m randomly distributed among them\cite{erdos_gallai,aigner,farkas,newman,frieze,berg,mizutaka}. All the different configurations of the graph have an equal probability to appear. We study the correlations for pairs of vertices by using the number of neighbors of each vertex, the so-called called degree $d(i)$, to represent the state of each vertex i.The emergence of correlations in uniform random graphs has been shown in several studies\cite{berg,mizutaka}. In our analysis the two vertices in each pair represent essentially the two spatially separated particles in Bell's experiment, while the degree represents the state of each particle. Then we form a linear combination of the respective joint probabilities that the two vertices in the pair have simultaneously specific degree values. We use the Wigner-d'Espagnat version of the Bell's inequality\cite{espagnat,bellderivWignerform} by defining the following quantity
\begin{equation}
B(a,b,c) = P(a,b) - P(a,c) -P(c,b),
\label{eq_bell}
\end{equation}
where $P(a,b)$ is the probability that one vertex in the pair has degree $d(i)=a$ and the other one has $d(i)=b$. If $B(a,b,c)>0$ then the Bell's inequalities are violated
implying the emergence of quantum mechanical correlations in the graph. We use the whole graph as a statistical ensemble to calculate the joint probabilities in Eq. \ref{eq_bell}. 

There are two mechanisms that lead to emergence of correlations between the degrees in the uniform random graph. Firstly there is the constraint that the number of edges in the graph is a fixed number m, so that the degrees have to satisfy the relation
\begin{equation}
\sum_{i=1}^{n} d(i)=2m.
\label{eq_constraint}
\end{equation}
Additional correlations are created due to structural constraints in the arrangement of the edges among the vertices. For example the degrees in Eq. \ref{eq_constraint} cannot take any value in the range $[1,m]$. The additional constraints can be expressed via the Erdos-Gallai theorem which states that the following 
\begin{equation} 
\sum_{i=1}^{k} d(i) \le k(k-1)+\sum_{i=k+1}^{n} min(d(i),k), 
\label{eq_constraint1}
\end{equation}
should hold for every k in $1 \le k \le n$ for simple graphs like the uniform random graphs considered in the current paper. We also note that the average degree in the graph is given by
\begin{equation}
d = \frac{2m}{n} \Rightarrow d=2R.
\label{eq2}
\end{equation}
where $R=\frac{m}{n}$ is the ratio of edges over vertices.

\begin{figure}
\begin{center}
\includegraphics[width=0.9\columnwidth,clip=true]{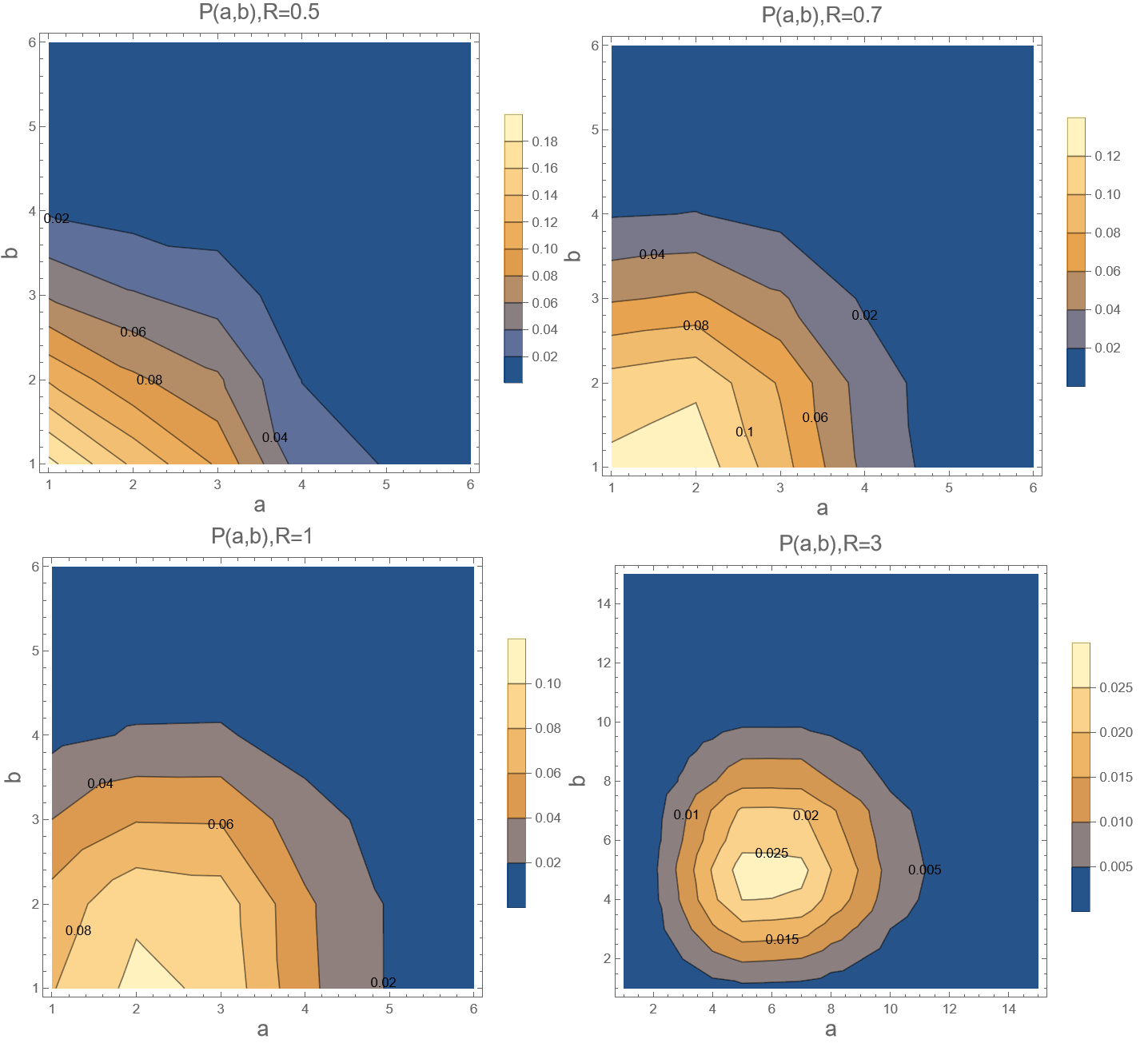}
\end{center}
\caption{The joint probability $P(a,b)$ for a pair of vertices with degrees a and b in the graph. The graph size is $n=3000$. For ratio of edges over vertices $R<2$ all the distributions are asymmetric around the average degree $d=2R$ indicating the existence of correlations between the degrees. For large ratio $R=3$ the distribution spreads normally around the average degree $d=6$ indicating the disappearance of the degree correlations.}
\label{fig1}
\end{figure}

The emergence of correlations between the degrees in the graph can be demonstrated by plotting the joint probability $P(a,b)$ for the vertex pair to have $a$ and $b$ degrees for its two vertices, the so called probability matrix, versus $a$ and $b$. In Fig. \ref{fig1} we show color plots of $P(a,b)$ versus $a,b$ for various ratios $R$. The asymmetrical distribution for $R \le 1$ signifies the appearance of correlations between the degrees. On the other hand, for $R=3$ the joint probability spreads symmetrically around the mean degree $d=2R=6$, signifying the disappearance of the degree correlations. The results in Fig. \ref{fig1} illustrate
that degree correlations emerge for sparse uniform random graphs for low $R$, near the region where the giant component in the graph appears at $R=0.5$. This means that although the constraints Eq. \ref{eq_constraint} and Eq. \ref{eq_constraint1} are always valid for every $R$, they lead to degree correlations only for small $R$.

\begin{figure}
\begin{center}
\includegraphics[width=0.9\columnwidth,clip=true]{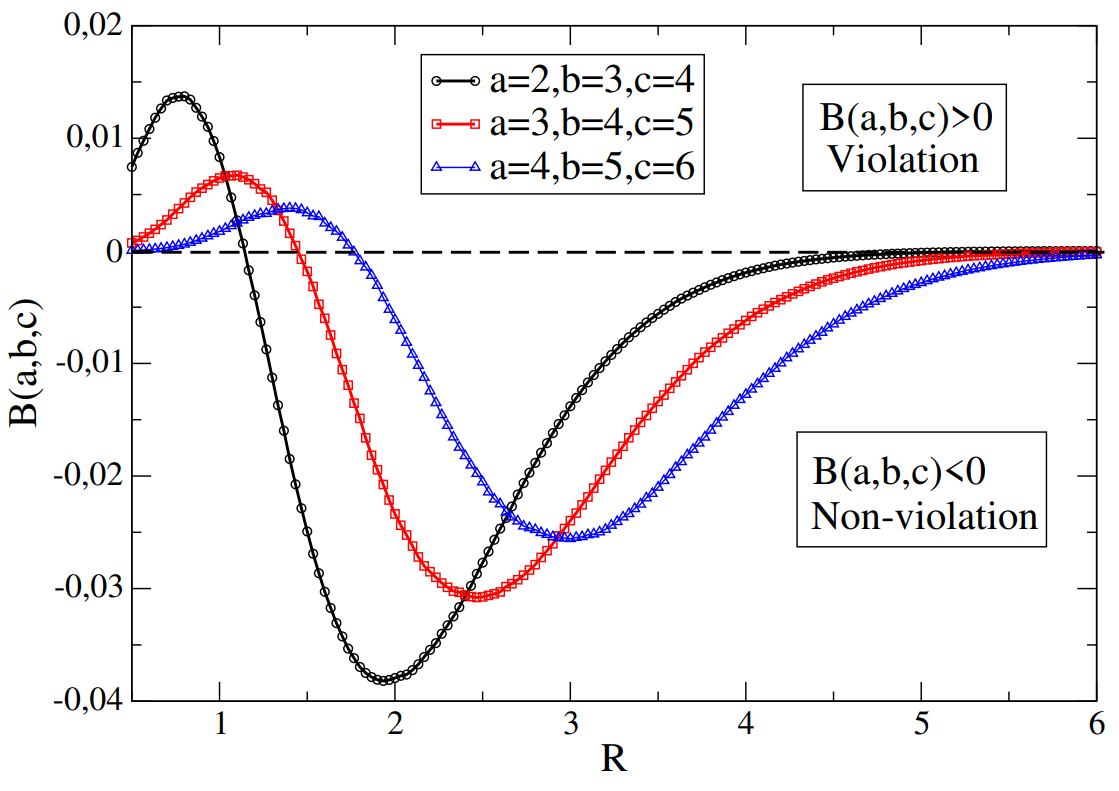}
\end{center}
\caption{The Bell's inequality $B(a,b,c)$ for various degree combinations $a,b,c$ and number of vertices $n=3000$. We have averaged over 100 runs(configurations) of the random graph. The inequality is violated for ratios $R<2$ indicating the emergence of quantum correlations in the graph.}
\label{fig2}
\end{figure}

For an Erdos-Renyi graph G(n,p), where p is the probability to have an edge, the probability for a vertex i to have degree $d(i)$ is given by the following formula\cite{frieze}
\begin{equation}
P(d(i)) = \binom{n}{d(i)} p^{d(i)} (1-p^{d(i) -1})^{(d(i))}.
\label{eq1}
\end{equation}
In addition when
\begin{equation}
p = \frac{2m}{n^{2}}
\label{eq3}
\end{equation}
graphs $G(n,p)$ and $G(n,m)$ behave similarly at the limit of a large number of vertices n. From Eq. \ref{eq2} and Eq. \ref{eq3} we get $p=\frac{d}{n}$.
For constant d and letting n go to infinity then Eq. \ref{eq1} becomes\cite{frieze,mizutaka}
\begin{equation}
P(d(i)) = \frac{(np)^{d(i)}}{d(i)!} e^{-np}=\frac{d^{d(i)}}{d(i)!} e^{-d} .
\label{eq4}
\end{equation}
We can define the Wigner-d'Espagnat version of the Bell's inequality as
\begin{equation}
P(d(i), d(i)+1) \leq  P(d(i), d(i)+2) + P(d(i)+2, d(i)+1),
\label{eq5}
\end{equation}
which corresponds to $a=d(i),b= d(i)+1,c=d(i)+2$ in Eq. \ref{eq_bell}.
At the limit of large ratio $R$, the correlations between the degrees become negligible as we have shown in Fig.\ref{fig1} and we can ignore them, assuming therefore that the degree probabilities of the vertices are statistically independent from each other.
For such random independent events, the joint probability in Eq. \ref{eq5} is just the product of the probability of each individual event $P(a,b) = P(a) P(b)$. Plugging the probabilities given by Eq. \ref{eq4} in Eq. \ref{eq5} we get
\begin{equation}
        d \leq  \frac{d^{2}}{(d(i)+2)} +  \frac{d^{3}}{(d(i)+2)(d(i)+1)}.
\label{eq6}
\end{equation}
Since $d=2R$, the above equation is always true for the $G(n,m)$ graphs when
\begin{equation}
R\geq\frac{1}{4}\left(-1-d(i)+ \sqrt{9+14d(i)+5d(i)^{2}}\right).
\label{eq8}
\end{equation}
Therefore when the degree correlations in the graph become negligible the Bell's inequality is satisfied for ratios $R$ given by Eq. \ref{eq8}.

\begin{figure}
\begin{center}
\includegraphics[width=0.9\columnwidth,clip=true]{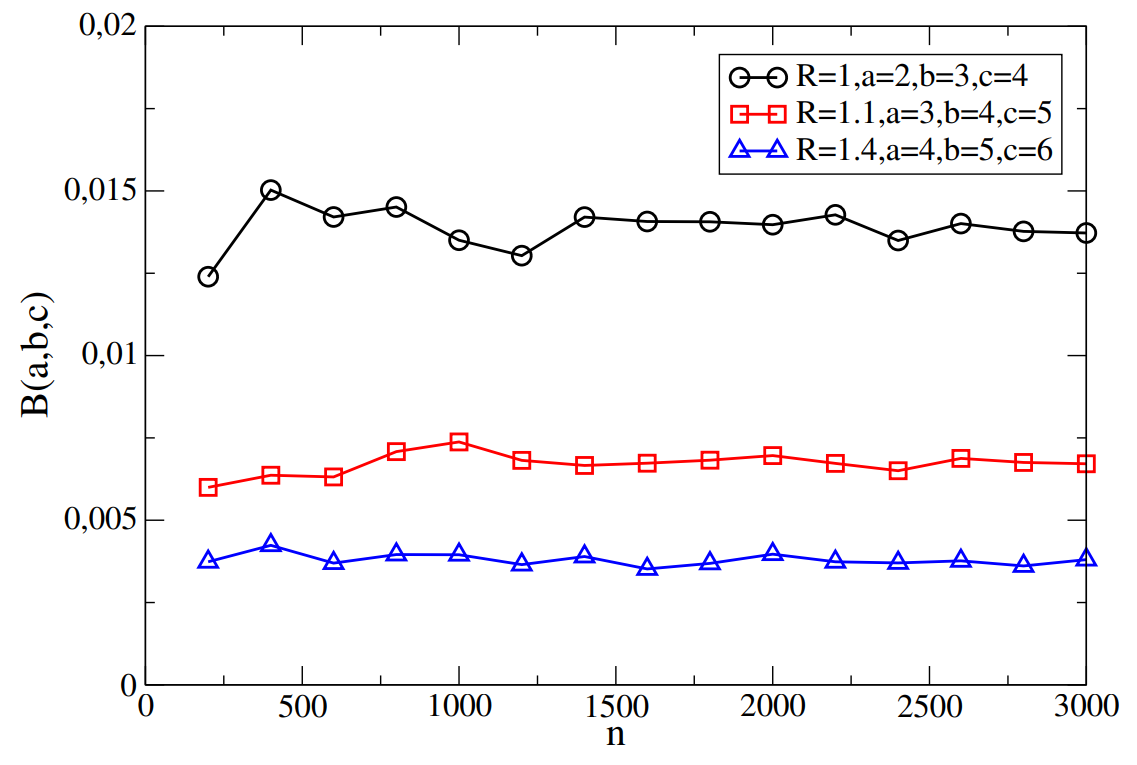}
\end{center}
\caption{The maximum value of the Bell's inequality $B(a,b,c)$ vs the number of vertices in the graph $n$, averaged over 100 runs. for large $n$ the strength of the correlations determined by the value of $B(a,b,c)$ converges for all degree combinations $a,b,c$.}
\label{fig3}
\end{figure}

In Fig. \ref{fig2} we plot the Bell's inequality $B(a,b,c)$ given by Eq. \ref{eq_bell}, for pairs of vertices in the graph, versus the ratio R, for graph size $n=3000$. Also, we average over 100 runs(configurations) of the graph. We have performed the calculation for three different degree combinations $a,b,c$. For sparse graphs in the region $R<2$ all degree combinations lead to $B(a,b,c)>0$. Consequently, the Bell's inequalities are violated in this region and quantum correlations between the vertices can emerge in the graph. In the region $R>2$, as the graph becomes denser by adding more edges between its vertices, all curves obtain negative values ($B(a,b,c)<0$). Therefore the Bell's inequalities are satisfied in this case, which also occurs when the degree correlations are not taken into account in our analysis, for values of $R$ given by Eq. \ref{eq8}. This result shows that a dense uniform random graph behaves as a classical statistical system with no quantum correlations between its vertices. We have found that $B(a,b,c)$ converges to the forms of the curves shown in Fig. \ref{fig2} for large $n$. The convergence is shown in Fig. \ref{fig3} where we plot the maximum value of $B(a,b,c)$ versus $n$ for the different degree combinations $a,b,c$. 

Consequently the Bell's inequalities are violated for sparse random graphs, with ratio of edges over vertices $R<2$, indicating the emergence of quantum correlations from the statistical correlations in the random graphs. The strength of the correlations becomes scale invariant for large graph sizes. 
It is important to remark that in one of our previous works, in the regime $R<2$ we have also found several emergent geometrical properties of the graph such as spatial dimension $D=3$ and curvature $K=0$, resembling the 3D space in models describing our universe\cite{paper1,paper2}. Combined with the results in the current paper, our overall analysis suggests that general geometrical properties like the dimensionality and the curvature along with quantum mechanical fluctuations can all emerge naturally in a random graph modeling of space depending on its ratio of edges over vertices, i.e. the spatial graph density. This result has fundamental implications concerning the emergence of spacetime, matter-energy and their quantum mechanical interactions from discrete models\cite{paper1,paper2,wolfram,gorard,markopoulou,trugenberger,trugenberger2}.

\begin{figure}
\begin{center}
\includegraphics[width=0.9\columnwidth,clip=true]{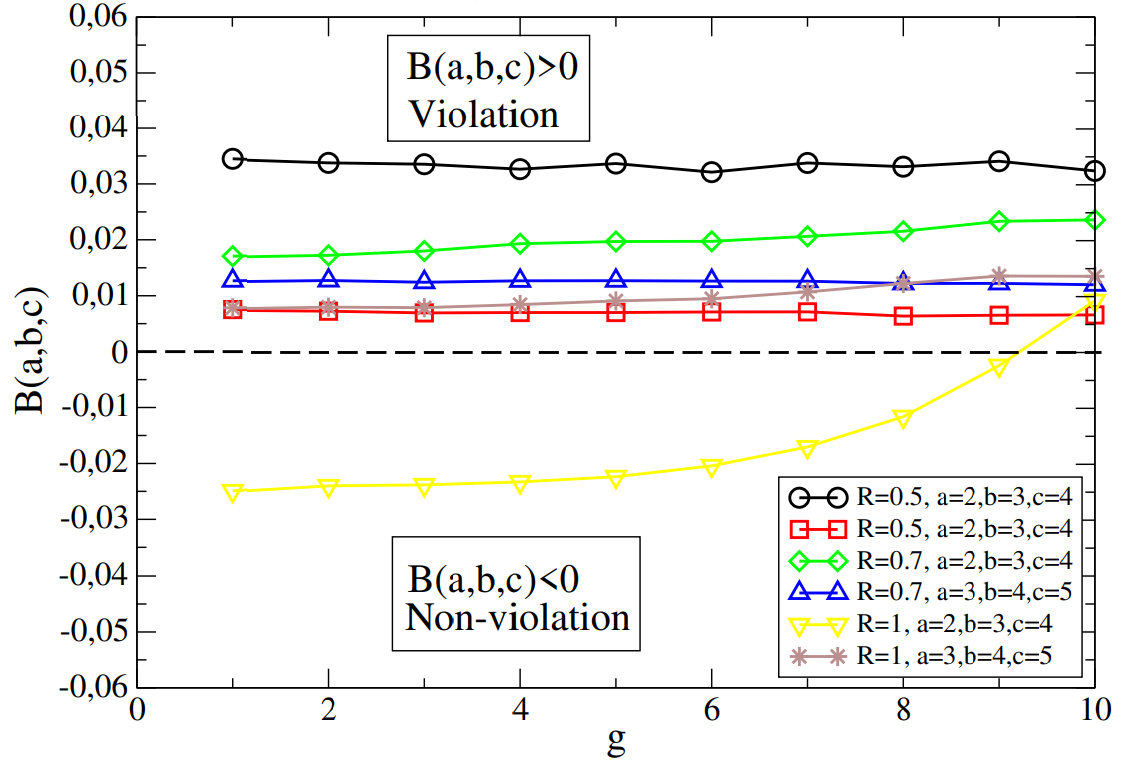}
\end{center}
\caption{The Bell's inequality $B(a,b,c)$ for pairs of vertices in the random graph vs the geodesic distance $g$ between the vertices, for various ratios $R$. We have averaged over 100 runs of the graph. The strength of the correlations determined by the value of $B(a,b,c)$ is independent of the geodesic distance between the vertices, indicating the non-local character of the correlations.}
\label{fig4}
\end{figure}

In Fig. \ref{fig4} we plot $B(a,b,c)$ for different geodesic distances $g$ between the vertices in the pair whose degree correlations we study, for different ratios $R$. The calculation is done for graph size is $n=3000$ and 100 runs. The geodesic distance is defined as the path between the two vertices that contains the minimum number of edges.  We can see that the value of $B(a,b,c)$ which characterizes the strength of the quantum correlations is relatively constant for all $R$ expect for $R=1$, independent of the geodesic distance between the correlated vertices. This results indicates the non-local character of the correlations, since every vertex in the graph could represent a spatial point in the D-dimensional manifold emerging out of the graph as we have shown in our previous works\cite{paper1,paper2}.

A fundamental proposition that arises from our analysis
is that the Bell's inequalities between two classical particles can be violated if correlations with other particles are taken into account statistically.
Consequently the inequalities for a pair of particles could be violated by considering the correlations induced by many other interacting particles acting as the bath of the system, which is usually ignored in experimental tests or theoretical calculations involving the Bell's inequalities. More generally our result strongly hints that quantum correlations can emerge in statistical physical models with many interacting components without assuming an a priori quantum mechanical nature of these models, like superpositions of different states of the system. One example of such statistical models is the uniform random graphs that we examine in the current paper, for large graph sizes by considering many vertices, and therefore many correlated degrees.

\begin{figure}
\begin{center}
\includegraphics[width=0.9\columnwidth,clip=true]{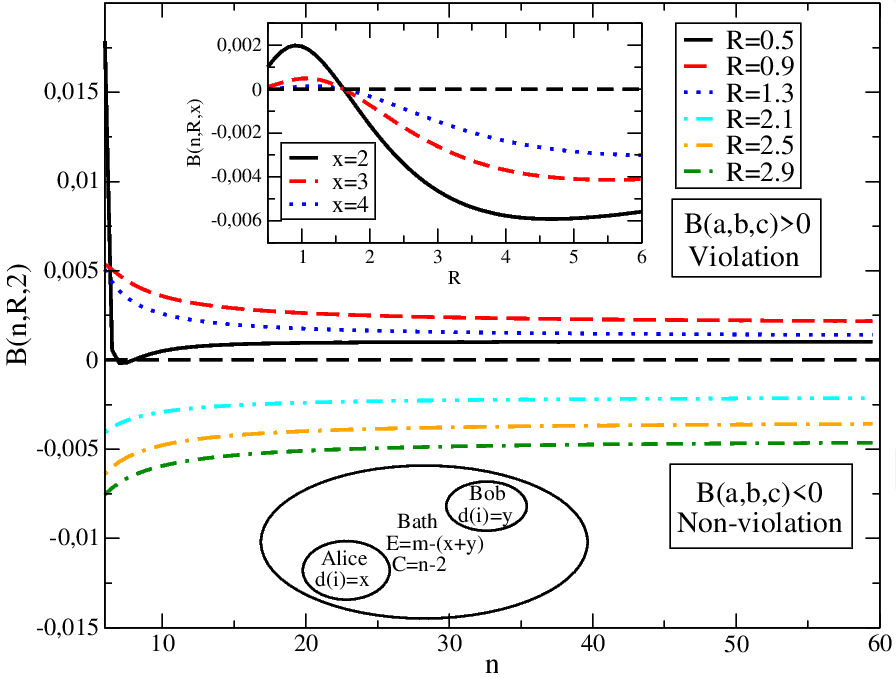}
\end{center}
\caption{Schematic: A toy model of a bosonic fluid of m bosons distributed among n states. The two entangled subsystems Alice and Bob contain x and y bosons respectively, leaving $m-(x+y)$ bosons to be distributed among $n-2$ states in the bath. Main figure: The Bell's inequality $B(n,R,x)$ between Alice and Bob versus the system size $n$ for different ratios $R=m/n$ and $x=2$. Inset: The Bell's inequality $B(n,R,x)$ vs $R$ for different $x$ at the limit $n \xrightarrow{} \infty$.}
\label{fig5}
\end{figure}

In order to probe the mechanism above we introduce a toy model of E bosons distributed among C states. The total number of configurations of the bosons among the states is given by
\begin{equation}
D(C,E)=\binom{C+E-1}{E}.
\label{eq_tm1}
\end{equation}
In the graph language the E bosons represent the m edges and the C states represent the number of vertices n in the graph. The occupation number of each state represents the degree of each vertex $d(i)$. This is equivalent to studying the distribution of the degrees among the vertices in the graph by considering only the constraint Eq. \ref{eq_constraint} and not Eq. \ref{eq_constraint1}. Then we can define the Bell's inequality between two states Alice and Bob, with boson occupation number $x$ and $y$ as shown in the schematic inside Fig \ref{fig5}. The remaining $m-(x+y)$ bosons are distributed among $n-2$ states forming the bath of the two states whose Bell's inequality we study. The number of configurations in the bath are given by Eq. \ref{eq_tm1} with $E=m-(x+y)$ and $C=n-2$. Then the Wigner-d'Espagnat inequality Eq. \ref{eq_bell} for $a=x,b=x+1,c=x+2$ becomes
\pagebreak

\begin{widetext}
\begin{equation}
B(n,m,x) = \frac{\binom{n-2+m-(2x+1)-1}{m-(2x+1)}-\binom{n-2+m-(2x+2)-1}{m-(2x+2)}-\binom{n-2+m-(2x+3)-1}{m-(2x+3)}}{\binom{n+m-1}{m}}.
\label{belltoy}
\end{equation}
\end{widetext}
In Fig. \ref{fig5} we plot Eq. \ref{belltoy} for $m=Rn$ and $x=2$. We can see that the Bell's inequality is violated for $R<2$ and is satisfied for $R>2$ as for the uniform random graph. In addition all curves saturate for large $n$. In the inset of Fig. \ref{fig5} we plot $B(n,m,x)$ at the limit $n\xrightarrow{} \infty$ versus $R$ for different $x$, given by
\begin{equation}
B(n,R,x) =\frac{R^{2x+1}}{(1+R)^{2x+3}}-\frac{R^{2x+2}}{(1+R)^{2x+4}}-\frac{R^{2x+3}}{(1+R)^{2x+5}}.
\label{eq_limit}
\end{equation}
The form of the curves is close to those in Fig. \ref{fig2}
although the order of magnitude of $B(n,m,x)$ is different. The above results show that the toy model, which takes account only the constraint Eq. \ref{eq_constraint} in the arrangement of the edges among the vertices, captures the core physical mechanism that leads to the violation of the Bell's inequalities for pairs of degrees in the uniform random graph.

To summarize we have demonstrated that the Bell's inequalities are violated for pairs of degrees in uniform random graphs that have a low ratio of edges over vertices, near the region where the giant component in the graph appears. Consequently the degree correlations in sparse uniform random graphs can mimic the correlations of quantum mechanical systems. More generally our result suggests that quantum mechanical correlations can emerge from the statistical correlations of random discrete models, without assuming a quantum mechanical nature of the random models on a fundamental level. This proposition has important implications concerning the emergence of spacetime, matter-energy and their quantum mechanical properties from random discrete models.

\section*{References}


\end{document}